# Equipartition of Current in Parallel Conductors on Cooling Through the Superconducting Transition

*S. Sarangi, S.P. Chockalingam, Raghav G Mavinkurve and S.V.Bhat**
*Department of Physics, Indian Institute of Science, Bangalore 560012, India*

*and*

*N.Kumar*
*Raman Research Institute, Bangalore 560080, India*

**Abstract**

Our experiments show that for two or more pieces of a wire, of different lengths in general, combined in parallel and connected to a dc source, the current ratio evolves towards unity as the combination is cooled to the superconducting transition temperature $T_c$, and remains pinned at that value below it. This re-distribution of the total current towards equipartition without external fine tuning is a surprise. It can be physically understood in terms of a mechanism that involves the flux-flow resistance associated with the transport current in a wire of type-II superconducting material. It is the fact that the flux-flow resistance increases with current that drives the current division towards equipartition.

In the course of our experiments with the original objective of settling the now not-so-frequently asked, but nevertheless askable question [1] as to whether the resistance in the superconducting state is absolutely zero, or merely too small, we were led to investigating how a given dc current divides between two or more conductors of different resistances connected in parallel as the combination is cooled through their common critical temperature $T_c$. To our surprise, the current ratio evolves monotonically, with cooling, towards unity, and stays pinned at that value below $T_c$. This equipartition of the total transport current between the two or more conducting arms connected in parallel, on approaching $T_c$, as also its persistence below $T_c$ does not follow from our conventional understanding of the macroscopic current transport through the normal or the superconducting circuits [2-4], to the best of our knowledge. After all, while the two resistances in parallel must go to zero simultaneously at their common transition temperature, their ratio could be arbitrarily different from unity.

The fact that the current ratio evolves from its usual Ohmic value (far from unity in the normal state much above $T_c$) to a "fixed point" value $\approx 1$ as T $\rightarrow T_c$, and remains pinned at this value in the superconducting state raises several basic questions, in addition to being of obvious relevance to superconducting circuits and electronics. Below we describe our experiment that uses contact-less current measurement (to avoid perturbing the current ratio). This is followed by a discussion of our main result, namely that of equipartition in physical terms of a mechanism involving the flux-flow resistances associated with the transport currents through the conductors connected in parallel.

The experimental arrangement used by us is sketched in fig.1. Here A and B form the two parallel arms of a superconducting loop of NbTi superconducting wire. A is a straight wire of length $l_A$ (about 6 cm in one of our experiments) and B, of length $l_B$, (~260 cm in our experiments) is in the form of a circular coil in the proximity of which a precision Hall probe sensor S is placed for contact-less current measurements. To start with, path A is disconnected and a current $I$ (= 6 Amperes in a typical experiment) from a constant current source is passed through path B. This generates a magnetic field $H_{n1}$ whose value (= 132 Gauss) is measured by the Hall probe Gauss meter. This contact-less probe ensures that the measurement itself does not perturb the current division in any way. Then the path A is connected in parallel with B (While doing this the respective ends of the two wire pieces are twisted together for about 2.5 cm each so that a single NbTi path connects the NbTi loop to the copper wire on either side. This is to eliminate the effects of any possible asymmetric bifurcation of the current and the consequent differential terminal resistance when the loop goes superconducting.). Now the field $H_{n2}$ generated by B is again measured. It is found that $H_{n2}$ = 3 Gauss which is nothing but $H_{n1}/(k+1)$, where $k = l_B / l_A$ is the ratio of the two wire lengths or equivalently of their resistances. This is of course, a simple consequence of the fact that the two resistances are connected in parallel.

Now the assembly is cooled to liquid helium temperature (4.2 K), i.e., below the transition temperature of the NbTi wire ($T_c$ = 9.3 K) and the magnetic field produced by the coil B is measured again. We find a value of 65 Gauss corresponding to a current flow of 3 A, i.e. one half of the total current. Figure 2 presents the results of this experiment along with the schematic diagram in the inset. In figure 3, the results of another experiment (carried out with a different coil) are also shown in which we had three parallel paths instead of two. In this case, we observed that one third of the total current passed through the coil. Thus the main observation of this work is that the current ratio evolves from its usual Ohmic value (far from unity in the present case) in the normal state much above $T_c$, to a value very close to unity as the transition (which had a small but finite width) was approached from above $T_c$. Moreover, the current ratio remained pinned at this value (i.e. close to unity) in the superconducting state below $T_c$. This evolution of the resistance ratio (or equivalently, the current ratio) converging to a "fixed point" ≈ 1, and its remaining pinned at this value below Tc seems to have escaped attention until now. This evolution of the current ratio towards equipartition in the transition region clearly involves some interesting physics --- of a re-distribution of the total current towards equality between the parallel arms without requiring any fine tuning. In the following a possible interpretation of this result is given in the physical terms of a mechanism involving the flux-flow resistance of a current carrying wire of the type-II superconducting material.

Consider first the flux-flow resistance of a type –II superconducting wire carrying a transport current. As is well known, the axial current generates an azimuthal magnetic field, and the corresponding flux is quantized into toroidal flux tubes linking the axial current. The latter exerts a Lorentz force causing the flux tubes to move radially. This radial motion of the flux tubes in turn induces an electric field along the axial transport current giving rise to dissipation, and hence equivalently to a resistance—the flux-flow resistance. This flux-flow resistance (dissipation), however, requires depinning of the flux tubes. Inasmuch as the depinning force is an increasing function of the transport current, the corresponding resistance increases with the latter. Thus, for two such conductors connected in parallel and carrying currents $I_1$ and $I_2$ with $I_1 + I_2 = I$, the impressed total current from a current source, we have $\partial R_1 / \partial I_1 > 0$, $\partial R_2 / \partial I_2 > 0$, implying a non-linearity, namely that the conductor initially carrying a higher current will have its resistance relatively raised, tending to suppress the current. This suppression provides a non-linear negative feedback that drives the currents in the two arms towards equality (equipartition) iteratively.

While the above flux-flow mechanism providing a resistive non-linear negative feedback is general, it is expected to iterate towards equipartition only for the experimental situation considered here, namely, where the system of conductors in parallel is already carrying the total impressed current *I* and is then cooled down towards the critical temperature, i.e. it is in the transition regime. (Indeed, the situation, where the system is already in the superconducting state, and then the current is impressed, is quite different. Here we should expect to observe the usual relation $I_1 L_1 = I_2 L_2$ to hold [2], where $L_1$ and $L_2$ are the inductances of the two arms. This indeed was verified by us to be the case.)

The reason for this can be appreciated from the following considerations. The transition region (T ≥ $T_c$) is dominated by the fluctuation superconductivity[5], where the pinning is relatively weak and the current induced depinning effectively controls the flux-flow resistance. Thus $\partial R_i / \partial I_i > 0$, (where 'i' labels the branches in parallel) together with the fact that $R_i \cong 0$ as T → $T_c$ drives the current division towards equipartition. Of course, once below $T_c$ the equipartition is protected through the superconductive diamagnetism (flux expulsion). A simple formal implementation of the above negative feedback due to the current dependent flux-flow resistance, supports the above physical picture. Indeed, it is interesting that a straightforward application of the extremal dissipation principle [6,7], i.e., extremising the total dissipation with respect to the transport currents $I_i$ subject to the constraint $\Sigma_i\ I_i = I =$ constant and $\partial R_i / \partial I_i > 0$ gives $\partial \ln R_1 / \partial \ln I_1 = \partial \ln R_2 / \partial \ln I_2 = \ldots\ldots\ldots$ The latter implies $I_1 = I_2 = \ldots\ldots$ in the limit $R_i \to 0$ ( i.e. approach to $T_c$ ) under the physical condition of monotonic current dependence of the logarithmic derivatives [8].

The negative non-linear feedback involving the transport-current induced pinning of the fluxons, and leading to equalization of the branch currents $I_1$ and $I_2$, can be derived more directly without depending on the minimum dissipation principle. The main point is that as the system is cooled through the transition region to the critical temperature, the wire resistances are increasingly dominated by the flux-flow resistances associated with the current-induced depinning, while the depinning itself becomes a sharply threshold-crossing process. We can then write $R_1(I_1) \equiv l_1\, r\,(I_1)$, $R_2(I_2) \equiv l_2\, r\,(I_2)$, where we have factored out the extensive 'length' parameters of the two wire pieces, and $r\,(I_{1,2})$ is an intrinsic resistive quantity (involving the current $I_{1,2}$ as the intensive variable), and controlled by the depinning as

$r\,(I_1) = r_o \exp\,[\,(I_1 - I_0)/\Delta\,]$, $r\,(I_2) = r_o \exp\,[\,(I_2 - I_0)/\Delta\,]$,

where $I_0$ is the threshold current, and $\Delta$ the width of the depinning 'threshold'.

Thus, for the two wire pieces in parallel, we have

$I_1 R_1(I_1) = I_2 R_2(I_2)$ with $I_1 + I_2 = I$, giving at once

$(I_1 - I_2)/I = (\Delta/I) \ln(l_2 I_2 / l_1 I_1)$. Now, as we approach the critical temperature, we expect $\Delta$ to become very small, and hence $I_1 \approx I_2$, with a difference which is logarithmic in the ratio of their lengths. Indeed, the equipartition is much stronger in as much as $I_1 l_1$ and $I_2 l_2$ are already approximately equal above the transition region.

Clearly, the interesting question as to how general the above conclusion is, requires further work, *e.g.,* trying different diameters and materials for the two resistances in parallel, but, of course, these must have the same critical temperature. An interesting system to study would be the case of wires of weak-linked granular materials, e.g. the high-$T_c$ superconducting ceramic materials, where the transport current induced suppression of the superconducting weak links in the wires is well described by the Lawrence-Doniach model [9], and should provide the non-linear negative feed back discussed above. Moreover, here the transition is much broader making it easier to probe the evolution towards equipartition experimentally.

In summary, we have reported a novel phenomenon with respect to the division of current between parallel conducting paths in which the total current is found to divide equally among the paths (irrespective of their inductances and the initial normal-state resistances) as the system is cooled to and below the common superconducting transition at $T_c$. This equipartition should be relevant to system geometries where the divided (series-parallel) superconducting circuit is cooled to and below Tc in the presence of a transport current. It also provides an interesting laboratory example for self-organization without fine tuning, and possibly for the application of the principle of extremal dissipation, that has been attracting considerable attention lately [6,7].

The funding provided by the University Grants Commission, India for this work is gratefully acknowledged. The authors thank S. Kasturirengan and K.V.Ramanathan for the loan of NbTi wire and the Gaussmeter respectively. One of us (NK) would like to thank the Department of Atomic Energy, (DAE), India, for the support received.

---------------------------------

*Electronic address for correspondence: svbhat@physics.iisc.ernet.in.

**FIGURE LEGENDS:**

Figure 1: The sketch of the experimental set up used. The part of the assembly enclosed in the dashed box could be inserted in an Oxford Instruments continuous flow cryostat for temperature variation. The coil B of diameter 20 mm has 40 turns of NbTi wire. Path A is made up of about 6 cm of straight length of the same NbTi wire. Respective ends of the two paths are twisted together for about 2.5 cm each so that a single NbTi path connects the NbTi loop to the copper wire on either side. This is to eliminate any possible effects of asymmetric bifurcation of the current and differential terminal resistance between the superconducting loop and the normal copper wire.

Figure 2: Equipartition in two parallel superconducting paths. The magnetic field measured and the corresponding current flowing through the coil are shown (a) when the entire current is flowing through path $P_2$ only and (b) when the path $P_1$ is connected in parallel with $P_2$. The inset shows the schematic of the experimental arrangement.

Figure 3: Equipartition in three parallel superconducting paths. The magnetic field measured and the corresponding current flowing through the coil are shown (a) when the entire current is flowing through the coil and (b) when two separate wires are connected in parallel with the coil.

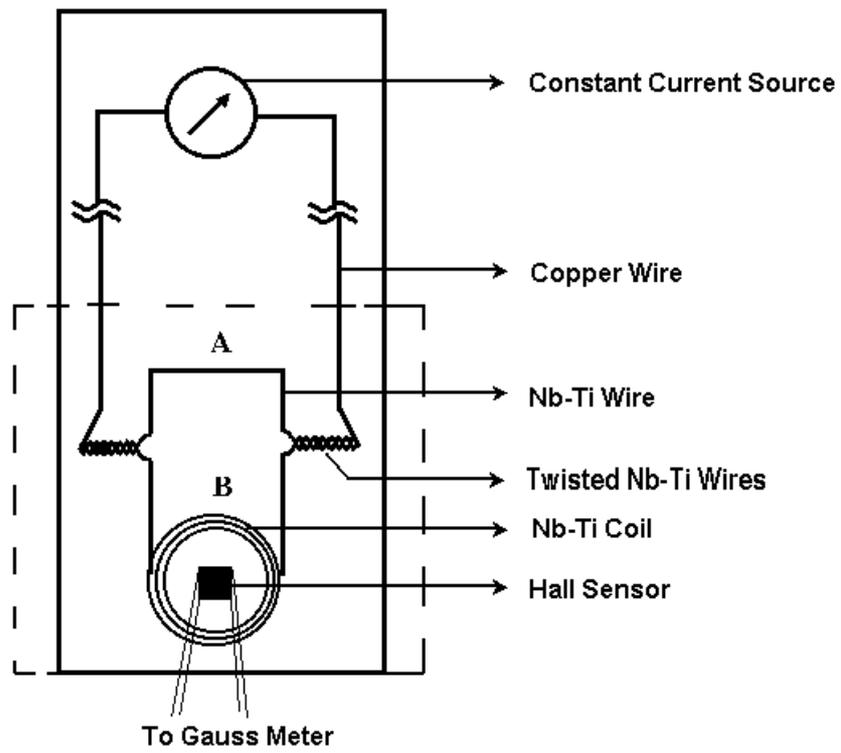

Figure 1: Sarangi et al.,

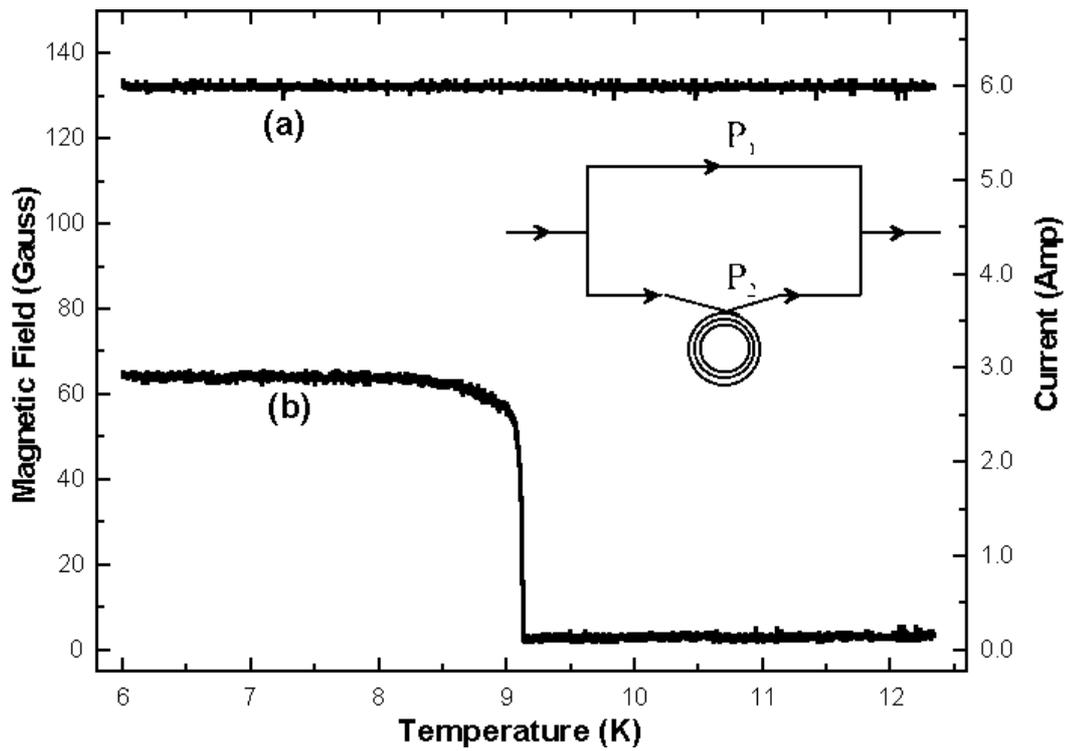

Figure 2: Sarangi et al.,

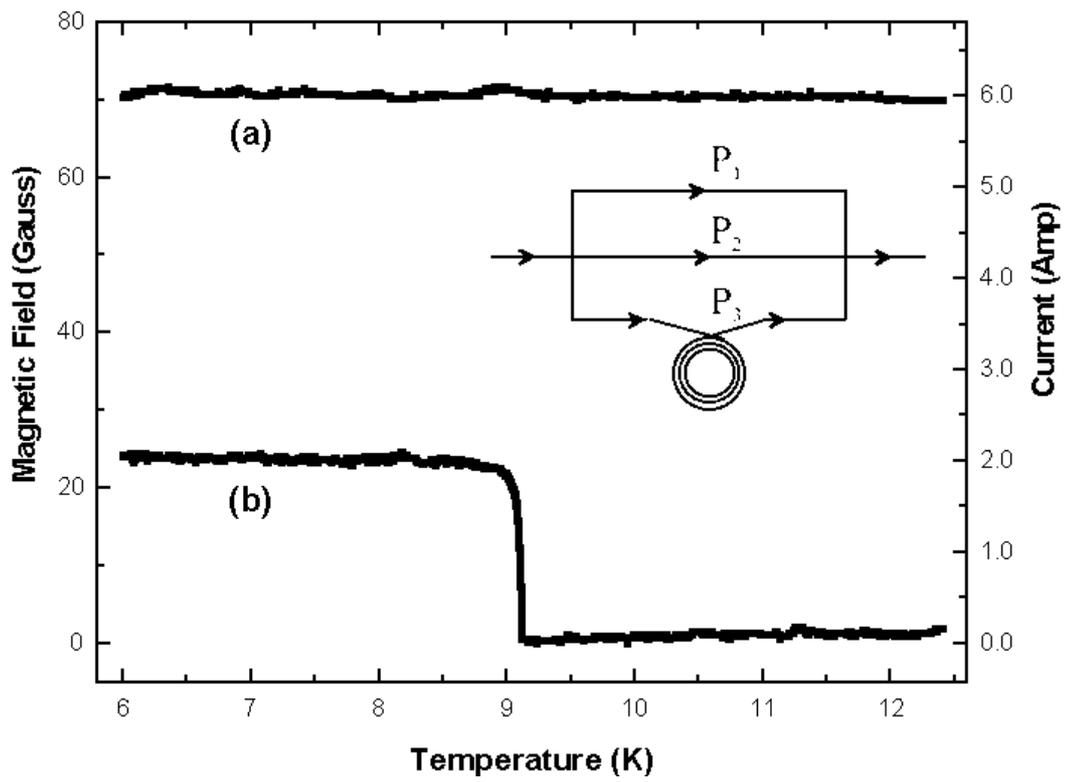

Figure 3: Sarangi et al,